\PassOptionsToPackage{svgnames}{xcolor}
\documentclass[twocolumn]{aastex702}
\usepackage{CJK}

\definecolor{linkcolor}{cmyk}{1,1,0,0}

\usepackage{color,soul}
\usepackage{scalerel}
\usepackage{dashrule}
\usepackage{totcount}
\newtotcounter{citenum}
\def\oldcite{}
\let\oldcite=\bibcite
\def\bibcite{\stepcounter{citenum}\oldcite}
\usepackage{mathtools}
\usepackage{tikz-cd}
\usepackage{wasysym}
\usepackage{amsmath}
\usepackage{fontawesome}

\usepackage[thinc]{esdiff}

\newcommand{\indep}{\perp \!\!\! \perp}
\newcommand{\dep}{\not\!\perp\!\!\!\perp}

\received{\today}
\submitjournal{AAS Journals}

\shorttitle{Hot Jupiter Inflation}
\shortauthors{Jin et al.}

\begin{document}
\begin{CJK*}{UTF8}{gbsn}

\author[0009-0000-2506-6645,gname=Zehao,sname=Jin]{Zehao Jin (金泽灏)}
\affiliation{Center for Astrophysics and Space Science (CASS), New York University Abu Dhabi, PO Box 129188, Abu Dhabi, UAE}
\affiliation{Center for Astronomy and Astrophysics and Department of Physics, Fudan University, Shanghai 200438, People's Republic of China}
\email[show]{\href{mailto:zj448@nyu.edu}{zj448@nyu.edu}}

\author[0000-0002-6633-376X,gname=Mohamad,sname=Ali-Dib]{Mohamad Ali-Dib}

\affiliation{Center for Astrophysics and Space Science (CASS), New York University Abu Dhabi, PO Box 129188, Abu Dhabi, UAE}
\email{malidib@nyu.edu}

\author[0009-0003-5225-6366,gname=Yujia,sname=Zheng]{Yujia Zheng (郑雨嘉)}
\affiliation{Carnegie Mellon University, PA, USA}
\email{yujiazh@cmu.edu}

\author[0000-0003-3784-5245,gname=Mario,sname=Pasquato]{Mario Pasquato}
\affiliation{INAF IASF-Milano, Via Alfonso Corti 12, 20133 Milano, Italy}
\affiliation{Ciela, Computation and Astrophysical Data Analysis Institute, Montreal, Quebec, Canada}
\email{mario.pasquato@inaf.it}

\author[0000-0002-4306-5950,gname=Benjamin,sname=Davis]{Benjamin L.\ Davis}
\affiliation{Center for Astrophysics and Space Science (CASS), New York University Abu Dhabi, PO Box 129188, Abu Dhabi, UAE}
\affiliation{Department of Physics, Astronomy, \& Materials Science, Missouri State University, Springfield, MO 65897, USA}
\email{ben.davis@nyu.edu}

\author[0009-0004-4596-7941,gname=Andrea,sname=Maccio]{Andrea V. Macci\`{o}}
\affiliation{Center for Astrophysics and Space Science (CASS), New York University Abu Dhabi, PO Box 129188, Abu Dhabi, UAE}
\email{maccio@nyu.edu}

\title{Causes of Hot Jupiter Inflation from Causal Discovery}
\begin{abstract}
Hot Jupiters often have radii larger than predicted by standard cooling--contraction models, but it remains unclear which process supplies or preserves the extra internal heat.
We analyze 328 short-period giant planets with measured \(M_p\), \(R_p\), \(P_{\rm orb}\), and host-star \(T_{\rm eff}\) using causal discovery, a statistical framework that asks which observed properties remain directly connected to planet radius after the others are accounted for.
As a check, the same pipeline recovers the expected mass--radius connection for a super-Earth control sample.
For hot Jupiters, the preferred graph links \(R_p\) directly to \(P_{\rm orb}\) and \(T_{\rm eff}\), but not to \(M_p\).
Since incident flux increases with \(T_{\rm eff}\) and decreases with \(P_{\rm orb}\) at fixed stellar properties, this paired dependence is naturally interpreted as a population-level signature of irradiation-regulated inflation.
Comparing the graph with analytic radius-excess scalings suggests a comparatively important role for Gold--Soter thermal tides, with kinetic/mechanical heating and ohmic dissipation potentially contributing alongside them.
Purely period-controlled gravitational tides are disfavored as the sole explanation because they lack a leading dependence on stellar temperature.
Distinguishing thermal tides, kinetic/mechanical heating, ohmic dissipation, and mixed scenarios will require radius-excess measurements that control for incident flux, age, composition, stellar properties, and selection effects.
More broadly, this work shows how causal discovery can turn population-level exoplanet data into physically interpretable tests of hot-Jupiter inflation.
Causal discovery complements parametric Bayesian population models by testing which observables retain direct conditional dependence on $R_p$ without imposing a specific radius relation, although the modest sample size limits the scope of the inferred graph.
\end{abstract}



\section{Introduction}\label{sec:intro}
\end{CJK*}
\linenumbers

Transiting hot Jupiters, gas giants on $P\lesssim10\,$d orbits, can show radii exceeding cooling--contraction predictions for their ages and compositions \citep{Charbonneau00,Burrows03,Fortney07,Laughlin11}.
Ground-based transit surveys such as HAT and WASP \citep{Bakos04HAT,Pollacco06WASP}, together with the space-based \textit{Kepler} and \textit{TESS} missions \citep{Koch10Kepler,Ricker15TESS}, have supplied a large population of transiting giant planets; analyses of these systems show significant inflation mainly above $F_{\rm inc}\gtrsim2\times10^8\,{\rm erg\,s^{-1}\,cm^{-2}}$, or $T_{\rm eq}\gtrsim1000\,$K for zero Bond albedo/full redistribution, with radius excess rising with $F_{\rm inc}$ and falling with planet mass \citep{Demory11,Weiss13,Sestovic18,Thorngren18}.

Long-lived inflation can result from cooling suppression, which delays planetary contraction, and/or from ongoing internal heating; genuine re-inflation as host stars brighten along the main or post-main sequence requires stellar-powered heat deposition capable of increasing the planetary interior entropy \citep{Lopez16,Grunblatt17,Komacek20,Thorngren21}.
Recent population modeling further indicates that hot-Jupiter cooling rates may be suppressed by approximately $95$--$98\%$, plausibly through shallow heating near or just below the radiative--convective boundary, thereby reducing the amount of deep heating required for re-inflation \citep{SchmidtThorngrenSchlaufman26}.

Models need not divide cleanly into cooling suppression and active heat deposition: deposited stellar power can both delay cooling and, when sufficiently deep or strong, drive re-inflation \citep{Ginzburg15,Komacek20,SchmidtThorngrenSchlaufman26}. 
Cooling suppression and active heat deposition are distinct processes that may coexist.
Shallow heating can efficiently delay contraction, but heating deposited above the radiative--convective boundary is generally ineffective at driving re-inflation \citep{Komacek20,SchmidtThorngrenSchlaufman26}.
Active channels include:
\begin{itemize}
    \item \textbf{ohmic dissipation} from ionized winds, efficient near $T_{\rm eq}\sim1500\,$K but weakened by low ionization or magnetohydrodynamic drag \citep{Batygin10,Menou12,Rogers14};
    \item \textbf{thermal tides} from stellar-heating-driven density perturbations that sustain asynchronous rotation and dissipate orbital/spin energy \citep{Arras09,Socrates13};
    \item \textbf{kinetic/mixing heating} by circulation, turbulence, or wave breaking at depth \citep{Youdin10,Sainsbury19}; and
    \item \textbf{gravitational tides} dissipated within the planet and driven by maintained eccentricity, planetary obliquity---the angle between the planet's rotation axis and its orbital angular-momentum axis---or asynchronous rotation, otherwise transient under circularization, obliquity damping, and synchronization \citep{Bodenheimer01,Jackson08,Millholland20_obliq}.
\end{itemize}

Population-level thermal-evolution analyses infer that the heating efficiency rises with irradiation, peaks at a few percent near $T_{\rm eq}\sim1500\,$K, and declines for hotter planets \citep{Thorngren18,Sarkis20}.
Distinguishing mechanisms requires constraints on wind speeds, magnetic fields, heat-deposition depths \citep{ThorngrenGaoFortney19}, and tidal responses via atmospheric characterization and structural probes such as Love numbers \citep{Kramm12}.

Causal discovery offers a complementary, data-driven route for this discrimination \citep{Spirtes2000}.
Rather than fitting one inflation model at a time, it asks which variables retain direct conditional dependence on the target observable after the remaining variables are accounted for.
Built on the foundation laid out by \citet{Spirtes2000} and \citet{pearl2009causality}, causal discovery has been widely used in epidemiology, genomics, condensed matter physics, and economics \citep{friedman2004inferring,sachs2005causal,runge2019inferring}.
In recent years, a number of applications in astrophysics are also emerging.
For example,  to infer causal structure in supermassive black-hole--galaxy coevolution \citep{Pasquato:2023,Jin2024beyond,Jin2025causalDiscoveryAstrophysics,Davis:2026}, to test whether trans-Neptunian-object colors are primordial \citep{Davis2025causalTNO}, to study the effect of environment on star formation \citep{mucesh2024}, to model stellar mass estimation \citep{zhang2025interpreting}, to decipher the stellar migration history in the Milky Way \citep{Jin2025archaeology}, and to explore the causal structure of galactic astrophysics \citep{Desmond2026causal}.

Here we apply the same logic to hot-Jupiter inflation: we compare the direct graph parents of \(R_p\) identified from the observational data with the expected signatures of delayed cooling, ohmic dissipation, thermal tides, kinetic/mechanical heating, and gravitational tides.
We start by deriving the radius-excess scalings used to translate each physical mechanism into predicted dependencies on $M_p$, $P_{\rm orb}$, and stellar $T_{\rm eff}$; we then describe the planet samples and causal-discovery pipeline, validate the approach on super-Earths, and interpret the hot-Jupiter graph in light of these theoretical scalings.

The remainder of this paper is organized as follows.
\S\ref{sec:methods} presents the analytic radius-excess scalings, planet samples, and causal-discovery pipeline.
\S\ref{sec:results} validates the method on a super-Earth control sample and applies it to the hot-Jupiter sample, interpreting the recovered graph in the context of delayed cooling, ohmic dissipation, thermal tides, kinetic/mechanical heating, and gravitational tides.
\S\ref{sec:summary} summarizes the implications for irradiation-regulated inflation and outlines the follow-up radius-excess tests needed to separate the remaining viable mechanisms.
Our Appendices provide added justifications of our dataset (\S\ref{app:robust}), clarify our adopted notation (\S\ref{app:notation}), give detailed theoretical derivations of the scaling relations (\S\ref{app:radius_scalings_details}), and explain further our sample selection and causal-discovery formalism (\S\ref{app:methods_details}).
The data and code used for this work are available for download from the following GitHub repository: \href{https://github.com/ZehaoJin/Causal-Hot-Jupiter-Inflation}{\faGithub~\url{https://github.com/ZehaoJin/Causal-Hot-Jupiter-Inflation}}.

\section{methods}\label{sec:methods}
\subsection{Analytic Radius-excess Scalings}\label{sec:radius_scalings}


We compare the causal-discovery graph to theoretical power-law scalings for the fractional radius excess, \(\Delta=(R_p-R_0)/R_0\), where \(R_0(M_p,t,Z)\) is the non-inflated radius at the same mass, age, and composition; rather than adopting a specific tabulated cooling track, we approximate its local mass dependence as \(R_0\propto M_p^{-0.06}\) \citep{Burrows03,Fortney07,Thorngren18}.

Here we show only the resulting dependence on \(M_p\), \(P_{\rm orb}\), and stellar \(T_{\rm eff}\); the full derivation, assumptions, and mechanism-by-mechanism table are moved to Appendix~\ref{app:radius_scalings_details}.
The scalings use \(F_{\rm inc}\propto T_{\rm eff}^{4}P_{\rm orb}^{-4/3}\), \(T_{\rm eq}\propto T_{\rm eff}P_{\rm orb}^{-1/3}\), \(R_0\propto M_p^{-0.06}\), \(T_{\rm int}\propto M_p^{0.50}\), and an active-heating response exponent \(\chi=0.20\).
For active heating, deposited power is converted to radius excess through \(\Delta_i\propto(P_i/L_{\rm int})^\chi\); delayed-cooling models are treated separately as low-irradiation benchmarks.
For compactness, the mechanism scalings used below are summarized as
\begin{equation}
\Delta_i \propto M_p^{\alpha_M}P_{\rm orb}^{\alpha_P}T_{\rm eff}^{\alpha_T},
\label{eq:Delta_powerlaw_main}
\end{equation}
with \((\alpha_M,\alpha_P,\alpha_T)\) listed in Table~\ref{tab:radius-theory-scaling}.

\begin{deluxetable*}{lccccc}
\centering
\tablecolumns{6}
\tablewidth{0pt}
\tabletypesize{\scriptsize}
\tablecaption{
Numerical Local Theoretical Scalings for the Radius-excess Law
$\Delta_i\propto
M_p^{\alpha_M}
P_{\rm orb}^{\alpha_P}
T_{\rm eff}^{\alpha_T}$.
\label{tab:radius-theory-scaling}
}
\tablehead{
\colhead{Scenario} &
\colhead{$\alpha_M$} &
\colhead{$\alpha_P$} &
\colhead{$\alpha_T$} &
\colhead{Relative Ordering of $|\alpha_j|$} &
\colhead{Qualitative Comparison with Recovered Graph}
}
\startdata
Ohmic dissipation, pre-turnover
& $-0.62$
& $-0.33$
& $+1.00$
& $|\alpha_T|>|\alpha_M|>|\alpha_P|$
& Leading $T_{\rm eff}$ term recovered; $M_p$ term absent
\\
Ohmic dissipation, near efficiency peak
& $-0.62$
& $-0.27$
& $+0.80$
& $|\alpha_T|>|\alpha_M|>|\alpha_P|$
& Leading $T_{\rm eff}$ term recovered; $M_p$ term absent
\\
Thermal tides, Gold--Soter
& $-0.42$
& $-0.60$
& $+0.60$
& $|\alpha_P|=|\alpha_T|>|\alpha_M|$
& Observed parents are co-leading; closest qualitative match
\\
Thermal tides, dynamical
& $-0.44$
& $-0.80$
& $\phantom{-}0.00$
& $|\alpha_P|>|\alpha_M|>|\alpha_T|=0$
& Period term recovered; observed $T_{\rm eff}$ parent not predicted
\\
Kinetic/mechanical heating
& $-0.40$
& $-0.27$
& $+0.80$
& $|\alpha_T|>|\alpha_M|>|\alpha_P|$
& $T_{\rm eff}$ recovered; $P_{\rm orb}$ is weaker than the absent
$M_p$ term
\\
Eccentricity/obliquity tides
& $-0.44$
& $-1.00$
& $\phantom{-}0.00$
& $|\alpha_P|>|\alpha_M|>|\alpha_T|=0$
& Period term recovered; observed $T_{\rm eff}$ parent not predicted
\\
Radiative opacity/delayed cooling
& $-0.78$
& $-0.08$
& $+0.25$
& $|\alpha_M|>|\alpha_T|>|\alpha_P|$
& Mass-dominated ordering and weak period term do not match
\\
Layered convection, fixed layer strength
& $-1.06$
& $\phantom{-}0.00$
& $\phantom{-}0.00$
& $|\alpha_M|>|\alpha_P|=|\alpha_T|=0$
& Only a mass dependence is predicted
\\
Potential-temperature advection
& $-1.06$
& $-0.33$
& $+1.00$
& $|\alpha_M|>|\alpha_T|>|\alpha_P|$
& Irradiation signature recovered; steep $M_p$ term absent
\\
\enddata
\tablecomments{
The entries are the exponents in Eq.~(\ref{eq:Delta_powerlaw_main}): $\alpha_M$ exponentiates $M_p$, $\alpha_P$ exponentiates $P_{\rm orb}$, and $\alpha_T$ exponentiates stellar $T_{\rm eff}$.
For this power-law form, the exponents are local logarithmic sensitivities, $\alpha_j=\partial\ln\Delta_i/\partial\ln X_j$, where $X_j\in\{M_p,P_{\rm orb},T_{\rm eff}\}$.
The fifth column reports the complete ordering of their absolute magnitudes and does not constitute a deterministic prediction of a causal parent set.
Every nonzero exponent represents a physical dependence of $\Delta_i$, even when the corresponding variable is not recovered as a parent of $R_p$.
Edge recovery additionally depends on the sampled ranges and covariances of the predictors, intrinsic and measurement scatter, and the distinction between the theoretically modeled radius excess $\Delta_i$ and the observed radius $R_p$.
The values assume $R_0\propto M_p^{-0.06}$, $T_{\rm int}\propto M_p^{0.50}$ at fixed age for active-heating channels, $\chi=0.20$ for active deep heating, fixed stellar mass and radius, fixed age and composition, fixed $B$, fixed $e$ or obliquity, and fixed tidal $Q'_p$.
The first ohmic row applies on the rising, pre-turnover branch; the second applies near the empirical heating-efficiency maximum.
}
\end{deluxetable*}

The key comparison is the relative ordering of the local theoretical sensitivities, rather than a deterministic mapping from exponent magnitude to a graph edge.
For the power-law form in Eq.~(\ref{eq:Delta_powerlaw_main}),
\begin{equation}
\alpha_j
=
\frac{\partial\ln\Delta_i}{\partial\ln X_j},
\qquad
X_j\in\{M_p,P_{\rm orb},T_{\rm eff}\}.
\label{eq:local_log_sensitivity}
\end{equation}
Consequently, every nonzero exponent represents a physical dependence of the radius excess.
Whether the corresponding variable is recovered as a parent of $R_p$ also depends on the sampled range and covariance of the predictors, intrinsic and measurement scatter, and the distinction between the modeled radius excess $\Delta_i$ and the observed radius $R_p$.

The mass exponent remains especially useful for distinguishing mechanisms whose inflation efficiency is strongly limited by higher surface gravity or faster intrinsic cooling from those whose leading observational signature is set primarily by irradiation.
Negative $\alpha_M$ means that, at fixed $P_{\rm orb}$ and $T_{\rm eff}$, lower-mass Jovian planets should show a larger fractional radius excess.
Values near $-0.4$ to $-0.6$ represent moderate, nonzero mass dependencies, whereas $\alpha_M\simeq-1$ implies a substantially stronger residual mass dependence.

Irradiation-dominated channels such as ohmic dissipation and kinetic/mechanical heating predict a strong positive $T_{\rm eff}$ dependence and a weaker negative period dependence.
Gold--Soter thermal tides \citep{GoldSoter69,Socrates13} predict $T_{\rm eff}$ and $P_{\rm orb}$ dependencies of equal and leading magnitude, together with a moderate mass dependence.
Purely gravitational eccentricity or obliquity tides predict a strong period dependence but no leading stellar-temperature dependence.
Delayed-cooling models serve as weak-irradiation-response benchmarks, although some have steep mass exponents because radius perturbations are suppressed by surface gravity. The numerical exponents used in \S\ref{sec:results} are listed in Table~\ref{tab:radius-theory-scaling}.

\subsection{Dataset} 

Our hot-Jupiter dataset was constructed from the \texttt{Encyclopaedia of exoplanetary systems}\footnote{\url{https://exoplanet.eu/catalog/}} \citep{Schneider11} by applying uniform cuts in planet mass, radius, and orbital period.
We adopted the operational hot-Jupiter criterion $P_{\rm orb}\leq 10\,\mathrm{days}$ used by \citet{Wright12}, rather than an irradiation- or equilibrium-temperature-based selection, and retained planets with $0.5\,\mathrm{M}_{\jupiter}\leq M_{\rm p}\leq 2.0\,\mathrm{M}_{\jupiter}$ and $0.5\,\mathrm{R}_{\jupiter}<R_{\rm p}\leq 3.0\,\mathrm{R}_{\jupiter}$.
The lower mass limit was adopted because the inflation behavior of planets below $0.5\,\mathrm{M}_{\jupiter}$ appears distinct and may be more strongly influenced by mass loss \citep{Thorngren18}.
More generally, these criteria select short-period, Jupiter-scale exoplanets while excluding objects with very small radii, extremely inflated radii, or masses outside the adopted giant-planet regime.

\begin{figure*}
\centering
\includegraphics[width=0.8\linewidth] {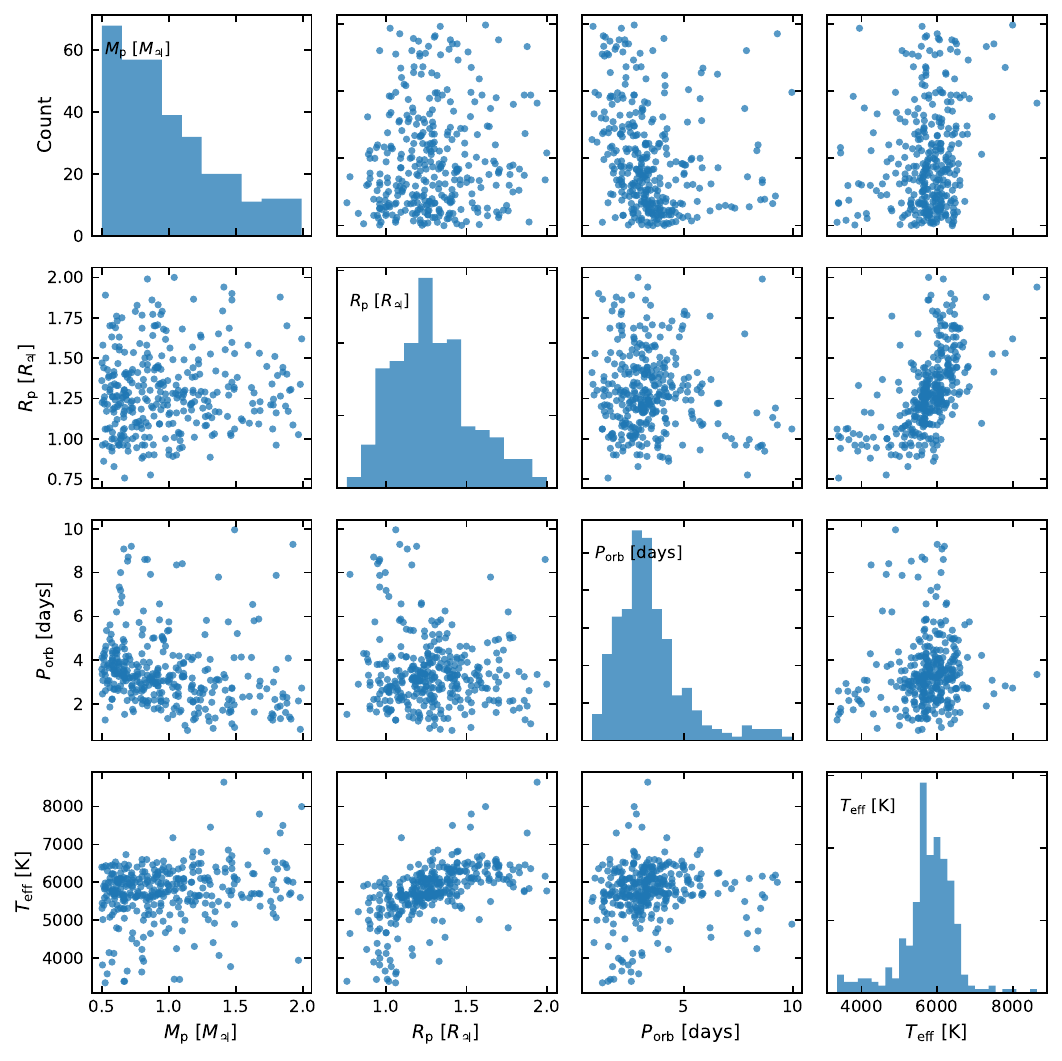}
\caption{Pair-plot matrix for the filtered hot-Jupiter sample, showing planet mass ($M_{\rm p}$), planet radius ($R_{\rm p}$), orbital period ($P_{\rm orb}$), and host-star effective temperature ($T_{\rm eff}$) for 328 exoplanets.}
\label{fig:pair-plot}
\end{figure*}

\begin{figure*}
\centering
\includegraphics[width=0.8\linewidth] {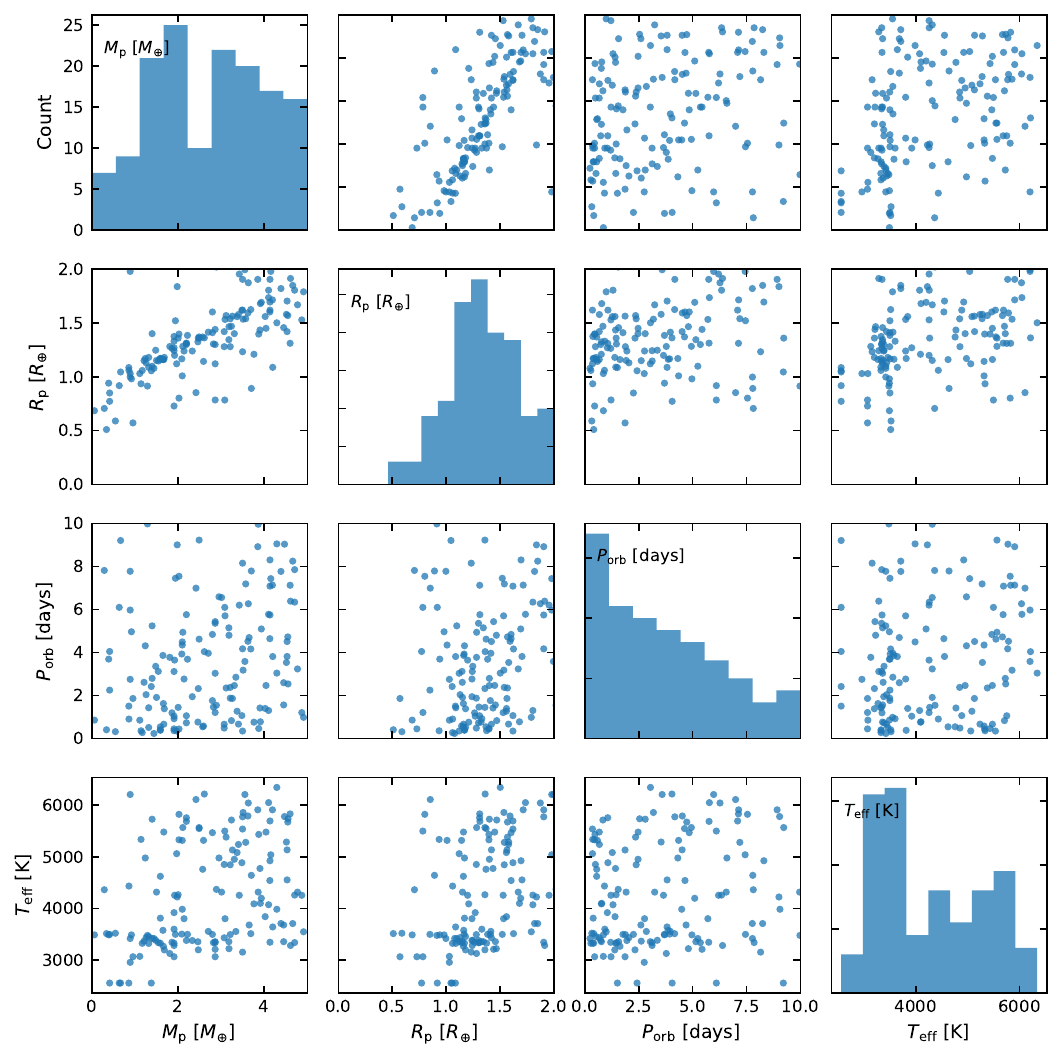}
\caption{Pair-plot matrix for the filtered super-Earth sample, showing planet mass ($M_{\rm p}$), planet radius ($R_{\rm p}$), orbital period ($P_{\rm orb}$), and host-star effective temperature ($T_{\rm eff}$) for 179 exoplanets.}
\label{fig:pair-plot2}
\end{figure*}

Because the cuts were implemented through direct numerical comparisons, planets lacking reported values for any of the filtered or analyzed quantities were excluded.
Requiring complete measurements of $M_{\rm p}$, $R_{\rm p}$, $P_{\rm orb}$, and $T_{\rm eff}$ yielded a final sample of 328 planets (Fig.~\ref{fig:pair-plot}).
As a sanity-check control sample, we also constructed a dataset of 179 transiting super-Earths with measured masses and radii from the same catalog, adopting $R_{\rm p}\leq 1.8\,\mathrm{R}_{\oplus}$ and requiring complete measurements of $M_{\rm p}$, $R_{\rm p}$, $P_{\rm orb}$, and $T_{\rm eff}$ (Fig.~\ref{fig:pair-plot2}).
No additional causal-discovery variables were used in the control analysis.

\subsection{Causal Discovery}
Uncovering causal structures from passively observed data is possible since different causal structures lead to distinct sets of conditional independence relationships among data.
For example, $A\leftarrow C\rightarrow B$ implies $A \dep B$ and $A \indep B\ |\ C$\footnote{``$A\rightarrow C$'' denotes ``$A$ causes $C$'', or $A$ is the causal parent of $C$. ``$A \dep B$'' reads as ``$A$ is dependent on $B$,'' and ``$A \indep B\ |\ C$'' reads as ``$A$ is independent of $B$ when conditioned on $C$.''}, but $A\rightarrow C\leftarrow B$ gives $A \indep B$ and $A \dep B \ |\ C$.
Therefore, one can go over possible causal graphs made by variables of interest, checking the likelihood that the observed data is compatible with the conditional independence relationships implied by the causal graph, until arriving at a causal graph with the highest likelihood.

We infer the causal structure among \(M_{\rm p}\), \(R_{\rm p}\), \(P_{\rm orb}\), and \(T_{\rm eff}\) using a search strategy called the Best Order Score Search \citep[BOSS;][]{andrews2023fast} with a likelihood defined by the Fourier Feature Marginal Likelihood (FFML) score \citep{ramsey2026fourier}.
All four variables were continuous and standardized before kernel scoring, making the analysis insensitive to physical units.

In brief, BOSS searches over variable orderings and projects each ordering to a directed acyclic graph (DAG, a graphical representation of causal structures) by adding or removing candidate parents when doing so improves the decomposable FFML score.
FFML evaluates nonlinear conditional relationships by approximating an RBF-kernel Gaussian-process marginal likelihood with random Fourier features \citep{RasmussenWilliams06,RahimiRecht07}, allowing each candidate edge to be tested by its improvement to the child variable's conditional model.
Graph edges are interpreted as conditional dependencies, not complete causal histories.
The full dataset description, pair-plot figure, ordering search, grow-shrink parent selection, and FFML equations are given in Appendix~\ref{app:methods_details}.

Notably, we guide the causal discovery process with a physics-informed prior on the direction of causal relations. 
Specifically, we impose that the planetary properties $M_{\rm p}$, $R_{\rm p}$, and $P_{\rm orb}$ cannot be causal parents of the host-star properties $T_{\rm eff}$ and stellar age. 
The prior therefore constrains only physically implausible causal directions and does not impose a particular mechanism for planetary inflation. In practice, incorporating this prior reduces the space of admissible DAGs and prevents the score-based search from using statistically favored but physically implausible reverse edges to orient otherwise ambiguous relationships. 
By incorporating well-established physical directionality while leaving the remaining edges to be determined by the data, the prior improves the interpretability and robustness of the inferred causal structure without directly biasing the analysis toward any specific inflation mechanism.

\section{Results}\label{sec:results}

\subsection{Super-Earths baseline}

\begin{figure}
    \centering
    \includegraphics[clip=true, trim=0mm 30mm 0mm 15mm, width=0.95\linewidth]{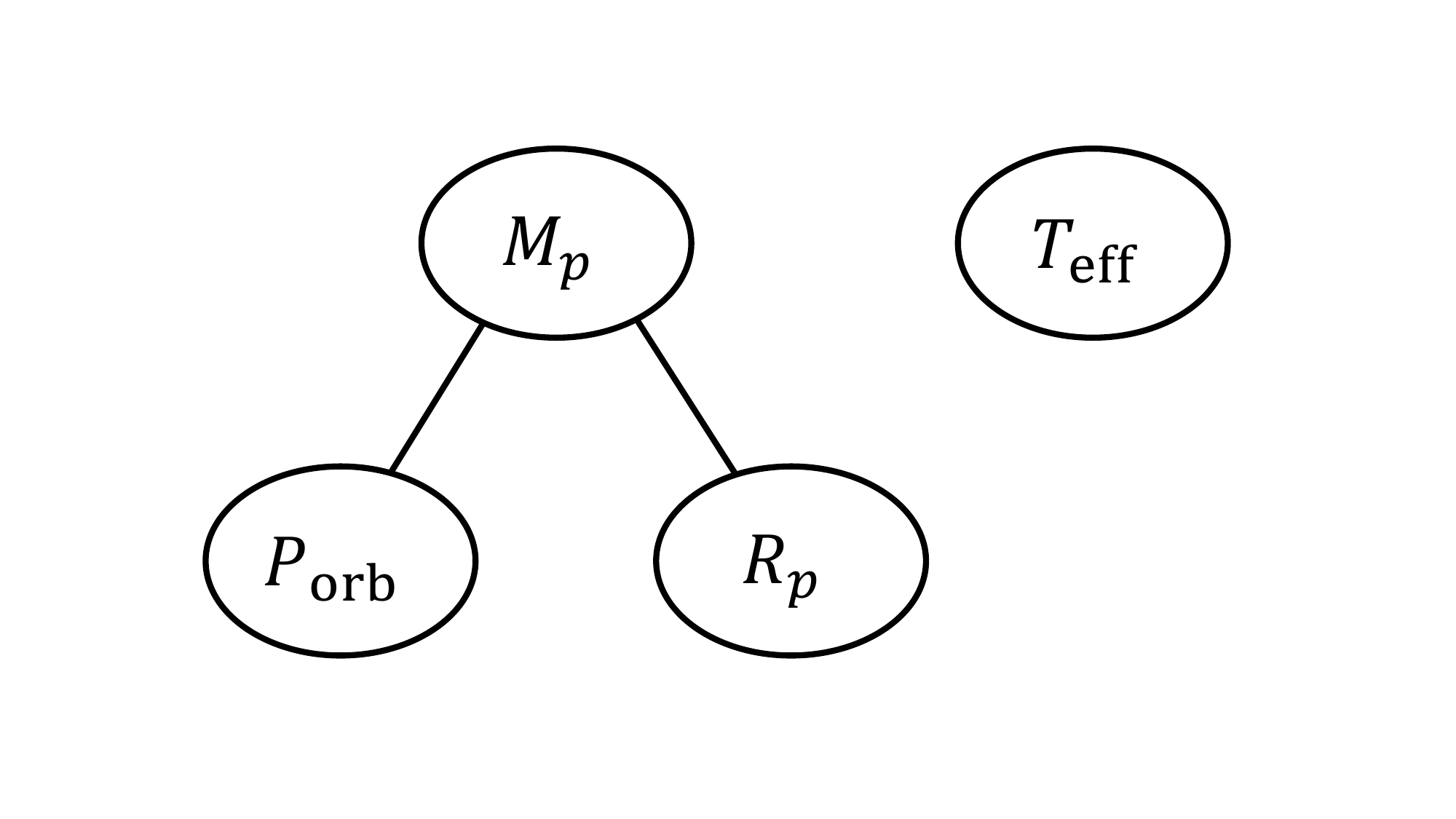}
    \caption{
    The causal structure found from super-Earth data using BOSS with FFML score. 
    An undirected edge $A\ - \ B$ suggests both directions are possible (either $A\rightarrow B$ or $A\leftarrow B$), as long as no new colliders are created.
    Here, the graph means that $P\rightarrow M_p\rightarrow R_p$, $P\leftarrow M_p\leftarrow R_p$, or $P\leftarrow M_p\rightarrow R_p$ (but \emph{not} $P\rightarrow M_p\leftarrow R_p$), plus an isolated node $T_\star$, are equally possible statistically and are together the most likely causal structure.
    }
    \label{fig1}
\end{figure}

As a sanity-check we first run our pipeline on the super-Earths dataset. Results are shown in Fig. \ref{fig1}. 
The recovered graph is consistent with the expected behavior of compact, mostly rocky or atmosphere-poor planets.
In this regime, \(R_p\) is expected to be governed primarily by mass and bulk composition, while irradiation mainly affects the probability of envelope loss rather than the radii of already stripped cores \citep{Seager07}.


The $M_{\rm p}$--$P_{\rm orb}$ edge may be driven largely by observational selection: radial-velocity measurements preferentially constrain more massive planets on shorter orbits, while transit-timing mass measurements depend strongly on system multiplicity, proximity to resonance, timing precision, and the observational baseline \citep{Steffen16,MillsMazeh17}.
Formation and migration may also contribute, but are not required to produce this edge.


The isolated $T_{\rm eff}$ node should not be interpreted as evidence that stellar properties or observational selection are unimportant.
Because transit depth scales as $\delta\simeq(R_{\rm p}/R_\star)^2$, correlations between $T_{\rm eff}$ and the omitted stellar radius $R_\star$ could affect the detectability of small planets.
The result only indicates that, within the selected sample and after conditioning on $M_{\rm p}$ and $P_{\rm orb}$, no additional direct $T_{\rm eff}$--$R_{\rm p}$ dependence is recovered.

Once planets have lost their primordial envelopes, stellar effective temperature no longer strongly controls their observed radii; irradiation matters mainly through atmospheric erosion \citep{OwenWu13}, not through the structure of bare rocky cores. 
With background knowledge above, we can further orient the DAG as: $P\leftarrow M_p\rightarrow R_p, T_\star$.

\subsection{Inflation of hot Jupiters}
\label{subsec:hot_jupiters_inflation}

\begin{figure}
    \centering
    \includegraphics[clip=true, trim=0mm 20mm 0mm 15mm, width=0.95\linewidth]{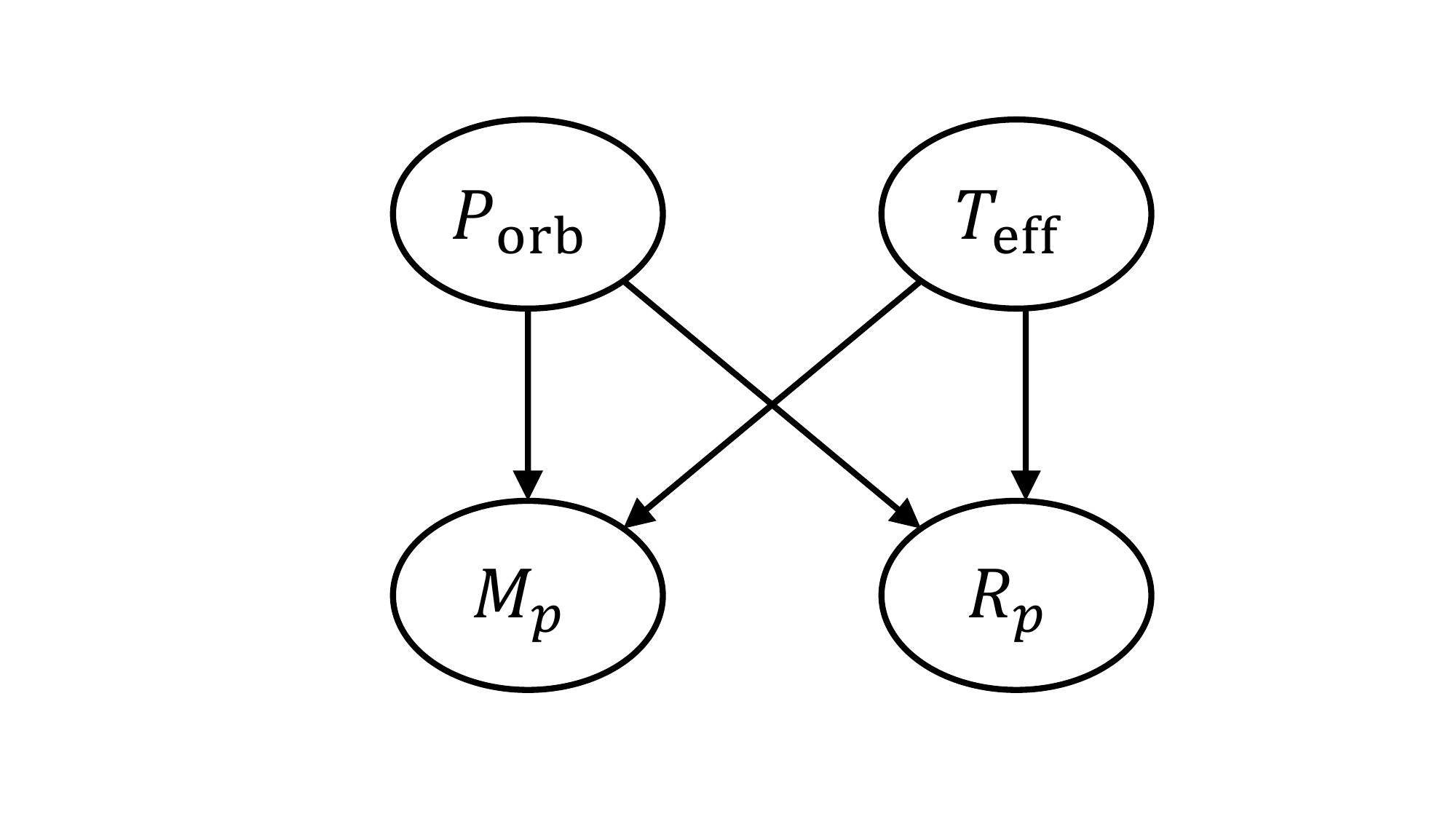}
    \caption{Preferred hot-Jupiter causal graph from BOSS with the FFML score. The inflation-relevant result is that $R_p$ has direct parents $P_{\rm orb}$ and $T_{\rm eff}$, but no direct $M_p\rightarrow R_p$ edge after conditioning.}
    \label{fig:hj_causal_graph}
\end{figure}

For the hot-Jupiter sample, the preferred four-variable graph in Fig.~\ref{fig:hj_causal_graph} contains
\begin{equation}
P_{\rm orb}\rightarrow M_p,\quad
T_{\rm eff}\rightarrow M_p,\quad
P_{\rm orb}\rightarrow R_p,\quad
T_{\rm eff}\rightarrow R_p .
\label{eq:final_dag_edges}
\end{equation}
Thus $R_p$ retains direct conditional dependence on orbital period and stellar effective temperature, but not on planet mass.
Because $F_{\rm inc}\propto T_{\rm eff}^{4}P_{\rm orb}^{-4/3}$ (under the fixed-\(M_\star\), fixed-\(R_\star\) scaling) is not an explicit graph variable, the paired $T_{\rm eff}\rightarrow R_p$ and $P_{\rm orb}\rightarrow R_p$ edges are the graph-level signature of irradiation-regulated inflation.
The arrows into $M_p$ should instead be read as population-level formation, migration, tidal-survival, photoevaporation/Roche-limit, and detection/selection effects \citep{FordRasio06,DawsonJohnson18}.

\begin{figure*}
    \centering
    \includegraphics[clip=true, trim=20mm 0mm 20mm 0mm, width=0.45\linewidth]{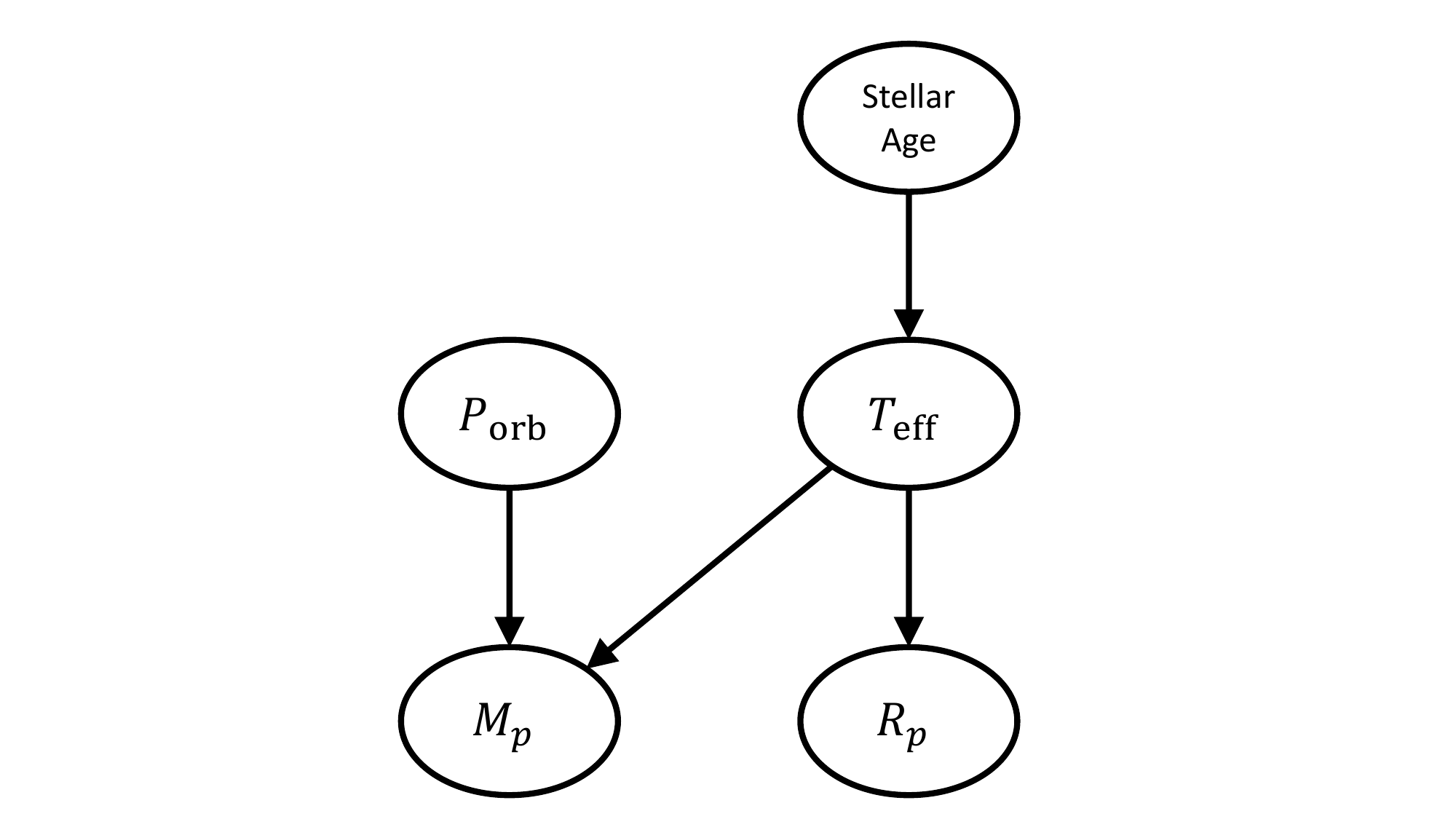}
    \includegraphics[clip=true, trim=20mm 20mm 20mm 15mm, width=0.45\linewidth]{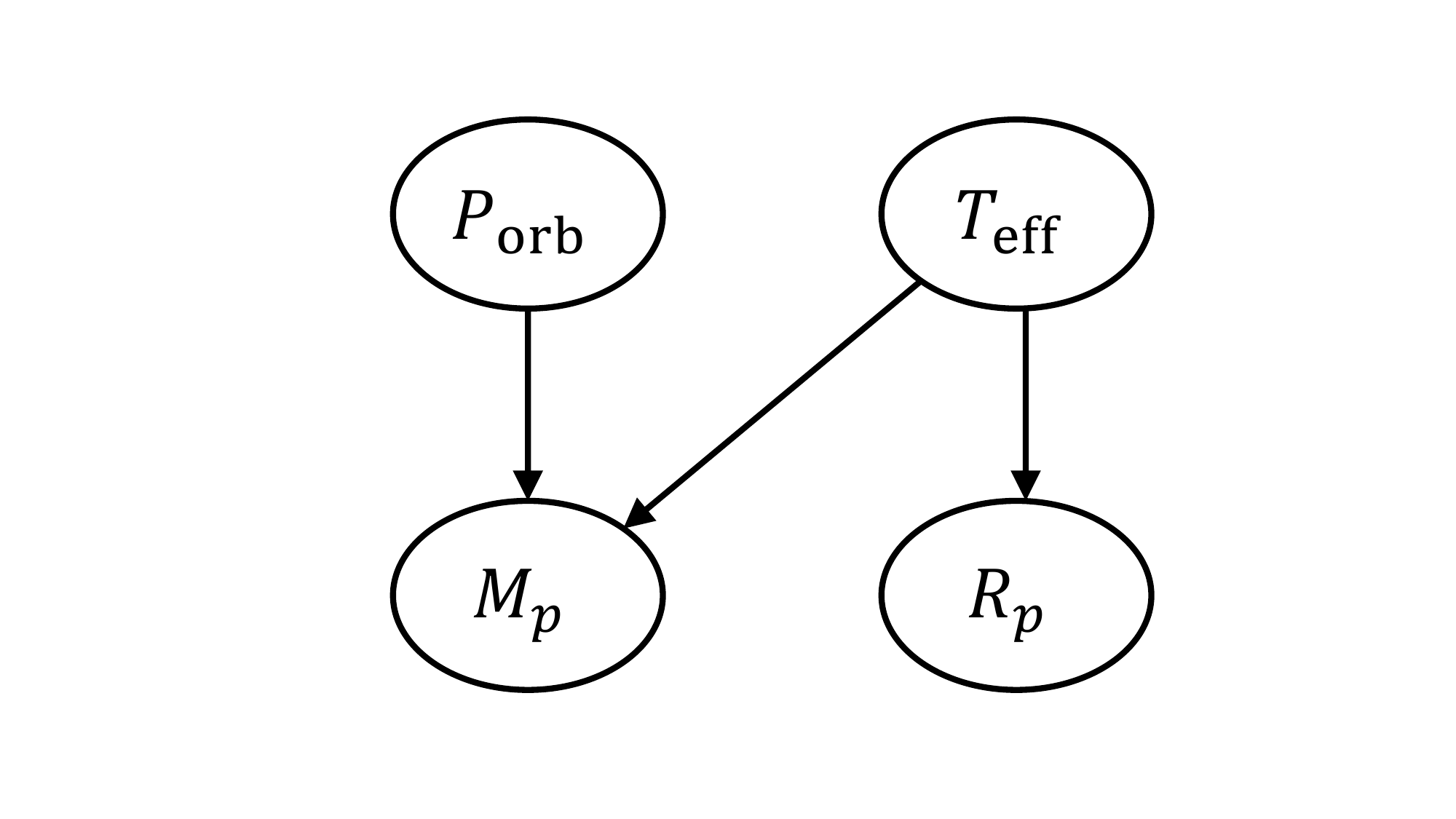}
    \caption{Left: Preferred hot-Jupiter causal graph from BOSS with the FFML score, with stellar age included. Right: same as left but without stellar age, using the same dataset reduced due to inclusion of age.}
    \label{fig:age}
\end{figure*}

Now we investigate if stellar age is a confounding variable in this analysis, given it determines the stellar effective temperature along with mass, composition, and rotation.
Stars' luminosities can vary by a factor of two or more across the main sequence, so its exclusion may theoretically affect the strength of our causal relationships.
In Figure~\ref{fig:age}, we show the preferred causal graph taking into account stellar age.
Including this variable implies a smaller dataset as age is not known for all stars in the original dataset.
Here we find the same graph, although with the $P_{\rm orb} \rightarrow R_p$ edge absent.
To highlight the origin of this discrepancy, we calculated the preferred causal graph using the same reduced dataset while excluding stellar age, and we find the same absence of $P_{\rm orb} \rightarrow R_p$, indicating that this is caused by the size of the dataset rather than an effect of stellar age.
Stellar age does determine stellar effective temperature as expected, but it does not directly cause $P_{\rm orb}, M_p$, and $R_p$, nor flip any existing causal edges.
We conclude that age is not a confounding variable in this analysis, and the irradiation-regulated inflation picture is still preferred with stellar age included.

Our preferred graph is not a measurement of the analytic exponents in Table~\ref{tab:radius-theory-scaling}: an edge indicates conditional dependence, not a local slope or a threshold in $|\alpha_j|$.
The absolute exponents instead rank the response of $\Delta$ to equal fractional changes in the predictors.
The missing $M_p\rightarrow R_p$ edge therefore does not eliminate a moderate mass scaling, because the graph is fitted to $R_p$ whereas the theoretical scalings describe $\Delta=(R_p-R_0)/R_0$.
Across the restricted Jovian mass range, the shallow non-inflated mass--radius relation, together with age, heavy-element content, entropy, opacity, and measurement scatter, can dilute a mass trend in $\Delta$.

For Gold--Soter thermal tides,
\begin{equation}
\Delta_{\rm th,GS}
\propto
M_p^{-0.42}
P_{\rm orb}^{-0.60}
T_{\rm eff}^{0.60},
\label{eq:Delta_thermal_GS_results}
\end{equation}
and therefore
\begin{equation}
|\alpha_P|
=
|\alpha_T|
=
0.60
>
|\alpha_M|
=
0.42.
\label{eq:Delta_thermal_GS_ordering}
\end{equation}
The two observationally selected parents of $R_p$, $P_{\rm orb}$ and $T_{\rm eff}$, are therefore the two co-leading theoretical sensitivities.
Gold--Soter thermal tides nevertheless predict a moderate residual mass dependence, whose absence from the recovered graph may reflect the use of $R_p$ rather than $\Delta$, the restricted mass range, and astrophysical scatter.

Kinetic/mechanical heating gives
\begin{equation}
\Delta_{\rm kin}
\propto
M_p^{-0.40}
P_{\rm orb}^{-0.27}
T_{\rm eff}^{0.80},
\label{eq:Delta_kinetic_results}
\end{equation}
with
\begin{equation}
|\alpha_T|
=
0.80
>
|\alpha_M|
=
0.40
>
|\alpha_P|
=
0.27.
\label{eq:Delta_kinetic_ordering}
\end{equation}
It therefore remains compatible with irradiation-regulated inflation, but the recovered graph contains its weakest theoretical dependence, $P_{\rm orb}$, while omitting the somewhat stronger $M_p$ dependence.
Ohmic heating, especially near the empirical efficiency peak, is also viable through Eqs.~(\ref{eq:Delta_ohmic_numeric}) and (\ref{eq:Delta_ohmic_peak_numeric}), although its stronger mass exponent makes the missing $M_p\rightarrow R_p$ edge less natural.

In this relative-ordering sense, Gold--Soter thermal tides provide the closest qualitative single-mechanism match to the recovered parent set, while kinetic/mechanical heating and ohmic dissipation remain viable.
The graph alone does not uniquely identify the underlying heating physics.
This conservative reading is consistent with the shallow-heating inference of \citet{SchmidtThorngrenSchlaufman26}, who find that hot-Jupiter cooling rates are suppressed by \(\simeq\)\(95\)--\(98\%\) and argue that heating near or just below the radiative--convective boundary can enable reinflation with much weaker deep heating.
We note that a flux-controlled radius excess with little residual \(P_{\rm orb}\) dependence would favor shallow ohmic dissipation or temperature advection, whereas a residual period term at fixed \(F_{\rm inc}\) would favor a thermal-tide contribution.

This parent set disfavors purely period-controlled gravitational tides as the sole population-level explanation.
Gold--Soter thermal tides are driven by stellar heating and predict both $T_{\rm eff}$ and period dependencies, whereas dynamical thermal tides and maintained eccentricity/obliquity tides, Eqs.~(\ref{eq:Delta_thermal_dyn_numeric}) and (\ref{eq:Delta_tides_numeric}), predict strong period dependencies but no leading $T_{\rm eff}$ dependence.
Delayed-cooling channels are not excluded, but radiative blanketing and fixed layered convection, Eqs.~(\ref{eq:Delta_radiative_numeric}) and (\ref{eq:Delta_layer_numeric}), have weak or absent direct irradiation/period dependencies.
Potential-temperature advection, Eq.~(\ref{eq:Delta_advection_numeric}), has the correct irradiation signature albeit with a steep mass exponent, making it more plausible as a contributor than as the cleanest single explanation.

The sharper discriminator is whether $\Delta$ retains residual $P_{\rm orb}$ dependence after specifying $F_{\rm inc}$.
Existing empirical radius--mass--flux relations and heating-efficiency estimates mostly use $F_{\rm inc}$ or $T_{\rm eq}$, rather than independent local slopes in $T_{\rm eff}$ and $P_{\rm orb}$ conditioned on age, composition, stellar properties, and selection effects \citep{Demory11,Weiss13,Thorngren18}.
Using Eq.~(\ref{eq:Finc_Teff_conversion}), irradiation-dominated mechanisms become
\[
\Delta_{\rm kin}\propto M_p^{-0.40}F_{\rm inc}^{0.20},
\qquad
\Delta_{\Omega,\beta_\Omega=1}\propto M_p^{-0.62}F_{\rm inc}^{0.20},
\]
with no leading residual period term, while Gold--Soter thermal tides become
\[
\Delta_{\rm th,GS}\propto
M_p^{-0.42}F_{\rm inc}^{0.15}P_{\rm orb}^{-0.40}.
\]
A residual $P_{\rm orb}$ dependence at fixed $F_{\rm inc}$ would favor Gold--Soter; a flux-controlled excess would favor kinetic/mechanical or ohmic heating.

Two-source models remain physically attractive.
A kinetic/mechanical plus Gold--Soter mixture,
\begin{equation}
\Delta_{\rm mix}=A_{\rm kin}\Delta_{\rm kin}+A_{\rm GS}\Delta_{\rm th,GS},
\label{eq:kinetic_gs_mixture}
\end{equation}
selects the observed parents with moderate mass exponents and interpolates between irradiation-dominated and balanced period--temperature slopes.
Locally,
\begin{equation}
\boldsymbol{\alpha}_{\rm eff}
\simeq
(1-f_{\rm GS})\boldsymbol{\alpha}_{\rm kin}
+
f_{\rm GS}\boldsymbol{\alpha}_{\rm th,GS},
\label{eq:mixture_slopes}
\end{equation}
with entries ordered as $(M_p,P_{\rm orb},T_{\rm eff})$, so increasing $f_{\rm GS}$ strengthens the period dependence with little change in mass dependence.
An ohmic plus Gold--Soter mixture,
\begin{equation}
\Delta_{\rm mix}=A_{\Omega}\Delta_{\Omega}+A_{\rm GS}\Delta_{\rm th,GS},
\label{eq:ohmic_gs_mixture}
\end{equation}
is plausible if many planets lie near the ohmic efficiency peak, but requires more dilution of the ohmic mass scaling by $R_p$-rather-than-$\Delta$ fitting, composition scatter, or age diversity.

At the present level, a single-source model is more parsimonious and a two-source model more flexible.
{The robust conclusion is that hot-Jupiter inflation is irradiation-regulated}; purely period-controlled gravitational tides are disfavored as the sole population-level channel; and Gold--Soter thermal tides, kinetic/mechanical heating, and ohmic dissipation remain viable.
The decisive follow-up is a radius-excess analysis in $(M_p,F_{\rm inc},P_{\rm orb})$, including stellar properties, age, metallicity, composition, and selection effects.
Variation between moderately irradiated and ultra-hot planets would point to mixed mechanisms rather than a universal source.

\section{Summary \& Conclusions}\label{sec:summary}

We applied nonlinear causal discovery to test which observed properties retain direct information about hot-Jupiter radii after conditioning on the others.
Using \(M_p\), \(R_p\), \(P_{\rm orb}\), and stellar \(T_{\rm eff}\), we compared the recovered graph with analytic scalings of the form
\[
\Delta\propto M_p^{\alpha_M}P_{\rm orb}^{\alpha_P}T_{\rm eff}^{\alpha_T},
\]
where \(\Delta=(R_p-R_0)/R_0\) is the fractional radius excess.
The main conclusions are:
\begin{itemize}
    \item As a control, the super-Earth sample recovers the expected mass--radius connection and leaves stellar \(T_{\rm eff}\) isolated, consistent with radii set mainly by interior structure once primordial envelopes are lost.

    \item For hot Jupiters, the preferred graph selects \(P_{\rm orb}\rightarrow R_p\) and \(T_{\rm eff}\rightarrow R_p\), but no direct \(M_p\rightarrow R_p\) edge after conditioning on the other variables.

    \item Because \(F_{\rm inc}\propto T_{\rm eff}^{4}P_{\rm orb}^{-4/3}\), the simultaneous dependence of \(R_p\) on \(T_{\rm eff}\) and \(P_{\rm orb}\) is naturally interpreted as evidence for irradiation-regulated inflation.

    \item The absence of a direct \(M_p\rightarrow R_p\) edge does not imply that mass is physically irrelevant.
    The graph is fit to \(R_p\), whereas the theory describes \(\Delta\); over the restricted Jovian mass range, composition, age, opacity, entropy, and measurement scatter can dilute moderate mass trends.

    \item Purely period-controlled gravitational tides are disfavored as the sole population-level explanation because they lack a leading stellar-temperature dependence.
    At the level of relative local-sensitivity ordering, Gold--Soter thermal tides provide the closest qualitative single-mechanism match: $P_{\rm orb}$ and $T_{\rm eff}$ are their co-leading theoretical sensitivities, while their residual mass dependence is moderate and nonzero.
    Kinetic/mechanical heating remains viable, but its period dependence is weaker than its unrecoverable mass dependence.
    Ohmic dissipation also remains viable, although its stronger mass scaling makes the missing $M_p\rightarrow R_p$ edge less straightforward.
\end{itemize}
The present four-variable graph constrains the parent set of \(R_p\), but not the residual-period behavior or local slopes of \(\Delta\).
Our ranking refers to the relative ordering of the theoretical sensitivities and should not be interpreted as evidence that Gold--Soter thermal tides predict no mass dependence.
A follow-up analysis should infer partial slopes for \(\Delta\) in \((M_p,F_{\rm inc},P_{\rm orb})\), including stellar properties, age, metallicity, composition, and selection effects.
A residual \(P_{\rm orb}\) dependence at fixed \(F_{\rm inc}\) would favor Gold--Soter tides; a mainly flux-controlled excess would favor kinetic/mechanical heating or ohmic dissipation.

\begin{acknowledgments}
This material is based upon work supported by Tamkeen under the NYU Abu Dhabi Research Institute grant CASS.
This research has made use of NASA's Astrophysics Data System.
This research was carried out on the high-performance computing resources at New York University Abu Dhabi.
The data and code used for this work are available for download from the following GitHub repository: \href{https://github.com/ZehaoJin/Causal-Hot-Jupiter-Inflation}{\faGithub~\url{https://github.com/ZehaoJin/Causal-Hot-Jupiter-Inflation}}.
\end{acknowledgments}


\software{
\href{https://causal-learn.readthedocs.io/en/latest/}{\textcolor{linkcolor}{\texttt{causal-learn}}} \citep{causallearn},
\href{https://github.com/matplotlib/matplotlib}{\textcolor{linkcolor}{\texttt{Matplotlib}}} \citep{Hunter:2007},
\href{https://networkx.org/}{\textcolor{linkcolor}{\texttt{NetworkX}}} \citep{networkx},
\href{https://github.com/numpy/numpy}{\textcolor{linkcolor}{\texttt{NumPy}}} \citep{harris2020array},
\href{https://pandas.pydata.org/}{\textcolor{linkcolor}{\texttt{Pandas}}} \citep{McKinney_2010},
\href{https://pgmpy.org/}{\textcolor{linkcolor}{\texttt{pgmpy}}} \citep{ankan2015pgmpy},
\href{https://pygraphviz.github.io/}{\textcolor{linkcolor}{\texttt{PyGraphviz}}},
\href{https://www.python.org/}{\textcolor{linkcolor}{\texttt{Python}}} \citep{Python},
\href{https://github.com/scipy/scipy}{\textcolor{linkcolor}{\texttt{SciPy}}} \citep{Virtanen_2020},
\href{https://seaborn.pydata.org/}{\textcolor{linkcolor}{\texttt{seaborn}}} \citep{Waskom2021}
}

\section*{ORCID iDs}

\begin{CJK*}{UTF8}{gbsn}
\begin{flushleft}
Zehao Jin (金泽灏) \scalerel*{\includegraphics{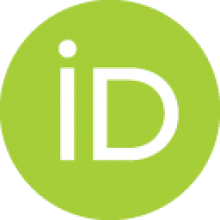}}{B}\\\url{https://orcid.org/0009-0000-2506-6645}\\
Mohamad Ali-Dib \scalerel*{\includegraphics{orcid-ID.png}}{B} \url{https://orcid.org/0000-0002-6633-376X}\\
Mario Pasquato \scalerel*{\includegraphics{orcid-ID.png}}{B} \url{https://orcid.org/0000-0003-3784-5245}\\
Yujia Zheng (郑雨嘉) \scalerel*{\includegraphics{orcid-ID.png}}{B} \url{https://orcid.org/0009-0003-5225-6366}\\
Benjamin L.\ Davis \scalerel*{\includegraphics{orcid-ID.png}}{B} \url{https://orcid.org/0000-0002-4306-5950}\\
Andrea V.\ Macci\`{o} \scalerel*{\includegraphics{orcid-ID.png}}{B} \url{https://orcid.org/0000-0002-8171-6507}
\end{flushleft}
\end{CJK*}

\appendix

\section{Dataset robustness}\label{app:robust}
As a catalog-sensitivity check, we applied the same nominal cuts and central-value completeness requirement to a July 2026 snapshot of the NASA Exoplanet Archive Planetary Systems Composite Parameters table.
The unrestricted selection contains 402 planets.
Because that table can include calculated radii for non-transiting systems, we also restricted the comparison to planets whose NASA discovery method is listed as Transit, yielding 376 planets, compared with 341 in the May 23, 2025 \url{https://exoplanet.eu/catalog/} snapshot before imposing the additional uncertainty-completeness requirement.
The two central-value selections share 318 planets, while 306 of the final 328 analysis planets occur in the NASA selection.
Their median values of \(M_{\rm p}\), \(R_{\rm p}\), \(P_{\rm orb}\), \(T_{\rm eff}\), and \(M_\star\) differ by at most \(2.2\%\).
Restricting both catalogs to discoveries through 2024 gives 337 NASA planets and 338 \url{https://exoplanet.eu/catalog/} planets, indicating that much of the numerical difference is attributable to the different catalog epochs.

\section{Notation}\label{app:notation}

Table~\ref{tab:notation} summarizes the symbols used in the analytic radius-excess calculations and causal-discovery formalism.
Symbols that have different meanings in different contexts are defined accordingly.

\begin{table*}[t]
\centering
\caption{Definitions of symbols used throughout the text.}
\label{tab:notation}
\scriptsize
\setlength{\tabcolsep}{5pt}
\renewcommand{\arraystretch}{1.08}

\begin{tabular}{@{}p{0.25\textwidth}p{0.69\textwidth}@{}}
\hline\hline
Symbol & Definition \\
\hline

\multicolumn{2}{@{}l}{\textit{Observed quantities and radius-excess scalings}}\\[1pt]

$M_p$, $R_p$, $P_{\rm orb}$, $T_{\rm eff}$
&
Planet mass, observed transit radius, orbital period, and host-star effective temperature, respectively.
\\

$P$, $T_\star$, $\delta$
&
$P$ denotes orbital period in the super-Earth graph discussion and atmospheric pressure in the ohmic-heating integral; $T_\star$ denotes the stellar-temperature node in the super-Earth graph; $\delta$ is the transit depth.
\\

$R_0(M_p,t,Z)$, $t$, $Z$
&
$R_0$ is the radius predicted by a non-inflated cooling track at planet mass $M_p$, age $t$, and composition $Z$.
\\

$\Delta$, $\Delta_i$, $\Delta R$
&
Fractional radius excess $\Delta=(R_p-R_0)/R_0$; $\Delta_i$ is the fractional radius excess associated with mechanism $i$; and $\Delta R=R_p-R_0$ is the dimensional radius excess.
\\

$F_{\rm inc}$, $T_{\rm eq}$
&
Incident stellar flux at the planet and planetary equilibrium temperature, respectively.
\\

$M_\star$, $R_\star$, $a$, $A_B$, $\sigma_{\rm SB}$
&
Host-star mass, host-star radius, orbital semimajor axis, Bond albedo, and Stefan--Boltzmann constant, respectively.
\\

$T_{\rm int}$, $L_{\rm int}$, $P_i$, $\chi$, $G$
&
Planetary intrinsic effective temperature, intrinsic cooling luminosity, power associated with heating mechanism $i$, local radius-response exponent, and gravitational constant, respectively.
The response is written as $\Delta_i\propto(P_i/L_{\rm int})^\chi$.
\\

$\alpha_M$, $\alpha_P$, $\alpha_T$
&
Exponents of $M_p$, $P_{\rm orb}$, and $T_{\rm eff}$ in $\Delta_i\propto M_p^{\alpha_M}P_{\rm orb}^{\alpha_P}T_{\rm eff}^{\alpha_T}$.
\\

$\xi_R$, $\xi_T$
&
Local mass exponents of the non-inflated radius and intrinsic temperature: $R_0\propto M_p^{\xi_R}$ and $\xi_T=\partial\ln T_{\rm int}/\partial\ln M_p$.
\\

$s_i$, $u_i$, $a_i$, $b_i$
&
Mechanism-dependent exponents in $P_i\propto R_0^{s_i}M_p^{u_i}F_{\rm inc}^{a_i}P_{\rm orb}^{-b_i}$.
\\

\hline
\multicolumn{2}{@{}l}{\textit{Heating, cooling, and circulation mechanisms}}\\[1pt]

$Q_{\rm ohm}$, $P_{\rm ohm}$, $P_\Omega$
&
Volumetric ohmic-dissipation rate, integrated ohmic power, and ohmic power used in the analytic scaling, respectively.
\\

$V_\phi$, $B$, $\eta$, $H_p$, $T$, $g$, $\beta_\Omega$
&
Zonal-wind speed, magnetic-field strength, magnetic diffusivity, pressure scale height, local atmospheric temperature, surface gravity, and local irradiation exponent of the ohmic power, respectively.
\\

$P_{\rm th,GS}$, $P_{\rm th,dyn}$,
$P_{\rm kin}$, $P_{\rm tide}$
&
Power associated with the Gold--Soter thermal-tide estimate, dynamical thermal tides, kinetic or mechanical heating, and maintained eccentricity or obliquity tides, respectively.
\\

$\epsilon_{\rm kin}$, $e$, $Q'_p$
&
Kinetic-heating efficiency, orbital eccentricity, and modified planetary tidal quality factor, respectively.
\\

$\Delta_\Omega$,
$\Delta_{\Omega,\,\beta_\Omega=1}$,
$\Delta_{\rm th,GS}$,
$\Delta_{\rm th,dyn}$
&
Radius-excess scalings for ohmic dissipation, ohmic dissipation at $\beta_\Omega=1$, Gold--Soter thermal tides, and dynamical thermal tides, respectively.
\\

$\Delta_{\rm kin}$, $\Delta_{\rm tide}$,
$\Delta_{\rm rad}$, $\Delta_{\rm layer}$, $\Delta_{\rm adv}$
&
Radius-excess scalings for kinetic or mechanical heating, maintained eccentricity or obliquity tides, radiative or opacity-delayed cooling, layered convection, and potential-temperature advection, respectively.
\\

$T_c$, $\delta T_{\rm deep}$
&
Characteristic convective-interior temperature and the irradiation-induced deep-atmosphere temperature perturbation, respectively.
\\

$\Delta_{\rm mix}$,
$A_{\rm kin}$, $A_{\rm GS}$, $A_\Omega$
&
Combined radius excess and the amplitudes of the kinetic, Gold--Soter, and ohmic contributions to the two-source models, respectively.
\\

$f_{\rm GS}$,
$\boldsymbol{\alpha}_{\rm eff}$,
$\boldsymbol{\alpha}_{\rm kin}$,
$\boldsymbol{\alpha}_{\rm th,GS}$
&
Local Gold--Soter fractional contribution, effective exponent vector, kinetic-heating exponent vector, and Gold--Soter exponent vector, respectively; vector entries are ordered as $(M_p,P_{\rm orb},T_{\rm eff})$.
\\

\hline
\multicolumn{2}{@{}l}{\textit{Causal discovery, BOSS, and FFML}}\\[1pt]

$A$, $B$, $C$, $\dep$, $\indep$
&
Generic variables and the dependence and conditional-independence relations used in the introductory causal-graph example.
In the grow--shrink procedure, $A$ is also the current parent set.
\\

$X_1,\ldots,X_p$, $p$
&
Variables entering the causal-discovery analysis and their total number; $p=4$ in the main analysis.
\\

$G$, $\operatorname{Pa}_G(j)$
&
Directed acyclic graph and the parent set of variable $X_j$ in graph $G$, respectively.
Here $G$ denotes the graph rather than the gravitational constant.
\\

$S(G)$, $S_{\rm FFML}$
&
Total decomposable score of graph $G$ and the local Fourier Feature Marginal Likelihood score, respectively.
\\

$\pi$, $\operatorname{Pre}_{\pi}(j)$
&
Candidate variable ordering and the variables preceding $X_j$ in that ordering, respectively.
\\

$A$, $X_A$, $k$
&
Current parent set, the variables indexed by that parent set, and a candidate parent index in the BOSS grow--shrink procedure.
In the FFML section, $k$ instead denotes the RBF kernel.
\\

$\Delta_{\rm add}(k;A,j)$,
$\Delta_{\rm del}(k;A,j)$,
$\widehat{\operatorname{Pa}}_{\pi}(j)$
&
Score gain from adding parent $k$, score gain from deleting parent $k$, and the final parent set selected for $X_j$ under ordering $\pi$, respectively.
\\

$Y$, $Z$, $n$, $d$
&
Child-variable data vector, candidate-parent data matrix, number of planets, and number of candidate parents, respectively.
Here $Z$ is a data matrix rather than the composition appearing in $R_0(M_p,t,Z)$.
\\

$f$, $\epsilon$, $\sigma^2$, $I_n$, $k$
&
Latent regression function, Gaussian noise term, noise variance, $n$-dimensional identity matrix, and RBF kernel, respectively.
\\

$K_Z$, $C_Z$, $S_{\rm KML}$
&
Kernel matrix for parent observations, covariance matrix $C_Z=K_Z+\sigma^2I_n$, and exact kernel marginal-likelihood score, respectively.
\\

$z$, $z'$, $z_i$, $h$
&
Inputs in candidate-parent space, the parent observation for planet $i$, and the RBF-kernel bandwidth.
\\

$\omega_\ell$, $b_\ell$, $\ell$, $m$
&
Random Fourier frequency, random phase, Fourier-feature index, and total number of random Fourier features, respectively.
\\

$\phi(z)$, $\Phi$
&
Random Fourier feature map and the feature matrix obtained by stacking $\phi(z_i)$ over the planets, respectively.
\\

$H$, $r$, $I_d$, $I_m$
&
Feature-space Gram matrix $H=\Phi^\top\Phi$, projected response $r=\Phi^\top Y$, and identity matrices of dimensions $d$ and $m$, respectively.
\\

\hline
\end{tabular}

\vspace{1mm}
\begin{minipage}{0.94\textwidth}
\scriptsize
\textit{Note.}
All symbols are reproduced exactly as used in the article.
When the same symbol has different meanings in different sections, both meanings are listed and distinguished by context.
Subscripts $i$, $j$, $k$, and $\ell$ index mechanisms, graph variables or candidate parents, and random Fourier features as indicated in the relevant equations.
Standard mathematical operators and conventional physical units are not listed separately.
\end{minipage}
\end{table*}

\section{Detailed Analytic Radius-excess Scalings}\label{app:radius_scalings_details}

The purpose of this section is to obtain explicit local power-law predictions for the fractional radius excess,
\begin{equation}
\Delta_i \equiv \frac{R_p-R_0}{R_0},
\label{eq:Delta_def}
\end{equation}
where $R_0(M_p,t,Z)$ is the radius expected from a non-inflated cooling track at the same mass, age, and composition.
The derived scalings describe the inflation term $\Delta_i$, not the absolute transit radius $R_p$.
Throughout this section $T_{\rm eff}$ denotes the stellar effective temperature.
The planet equilibrium temperature $T_{\rm eq}$ is used only as an intermediate irradiation variable; the final radius-excess scalings are written in terms of $M_p$, $P_{\rm orb}$, and stellar $T_{\rm eff}$.

The incident flux and equilibrium temperature are
\begin{equation}
\begin{split}
F_{\rm inc}
&=\sigma_{\rm SB}T_{\rm eff}^{4}
  \left(\frac{R_{\star}}{a}\right)^2
  \propto
  T_{\rm eff}^{4}R_{\star}^{2}M_{\star}^{-2/3}
  P_{\rm orb}^{-4/3},\\
T_{\rm eq}
&=\left[
  \frac{(1-A_B)F_{\rm inc}}
       {4\sigma_{\rm SB}}
  \right]^{1/4}
  \propto
  T_{\rm eff}R_{\star}^{1/2}M_{\star}^{-1/6}
  P_{\rm orb}^{-1/3}.
\end{split}
\label{eq:Finc_Teq_numeric}
\end{equation}
In the scalings below, we hold $M_{\star}$, $R_{\star}$, albedo, age, and composition fixed, or equivalently absorb the residual stellar-radius and stellar-mass factors into the normalization.
Thus the working conversion is
\begin{equation}
F_{\rm inc}\propto T_{\rm eff}^{4}P_{\rm orb}^{-4/3},
\qquad
T_{\rm eq}\propto T_{\rm eff}P_{\rm orb}^{-1/3}.
\label{eq:Finc_Teff_conversion}
\end{equation}

We now replace the free response parameters by representative numerical values.
For the uninflated giant-planet mass--radius slope we adopt
\begin{equation}
R_0\propto M_p^{\xi_R},
\qquad
\xi_R=-0.06,
\label{eq:xiR_numeric}
\end{equation}
consistent with the nearly flat mass--radius relation of Jovian planets over the hot-Jupiter mass range \citep{Burrows03,Thorngren18}.


For the intrinsic luminosity, we adopt a mechanism-independent, fixed-age Kelvin--Helmholtz scaling.
We do not use the irradiation-dependent $T_{\rm int}(T_{\rm eq})$ relation of \citet{ThorngrenGaoFortney19}, because it assumes equilibrium between empirically inferred anomalous heating and the outgoing intrinsic flux; using it in $P_i/L_{\rm int}$ would therefore insert a heating prescription into the baseline cooling luminosity and partly double-count the irradiation dependence.
With $R_0$ only weakly dependent on mass, $L_{\rm int}\sim GM_p^2/(R_0t)$ and $L_{\rm int}\sim R_0^2T_{\rm int}^4$, giving $T_{\rm int}\propto M_p^{1/2}$ at fixed age.
Thus we adopt
\begin{equation}
\xi_T\equiv
\frac{\partial\ln T_{\rm int}}
     {\partial\ln M_p}
=0.50.
\label{eq:xiT_numeric}
\end{equation}
For active deep heating, we adopt
\begin{equation}
\chi=0.20,
\label{eq}
\end{equation}
motivated by the analytic deep-deposition response exponent $\delta\simeq0.19$ derived by \citet{Ginzburg15} for representative opacity laws and its comparison with numerical evolution calculations by \citet{KomacekYoudin17}.
We treat $\chi$ as a local, order-of-magnitude response exponent rather than a universal constant.

For any active heating mechanism with
\begin{equation}
P_i\propto R_0^{s_i}M_p^{u_i}F_{\rm inc}^{a_i}P_{\rm orb}^{-b_i},
\label{eq:generic_active_numeric_power}
\end{equation}
the local cooling-time-consistent response is
\begin{equation}
\Delta_i
\propto
\left(\frac{P_i}{L_{\rm int}}\right)^{\chi}
\simeq
\left(\frac{P_i}{R_0^2T_{\rm int}^{4}}\right)^{\chi}.
\label{eq:active_response_numeric}
\end{equation}
Using Eq.~(\ref{eq:Finc_Teff_conversion}), this gives
\begin{equation}
\Delta_i
\propto
M_p^{\left[u_i+(s_i-2)\xi_R-4\xi_T\right]\chi}
T_{\rm eff}^{4a_i\chi}
P_{\rm orb}^{-\left(b_i+4a_i/3\right)\chi}
\propto
M_p^{0.20\left[u_i-0.06(s_i-2)-2.00\right]}
T_{\rm eff}^{0.80a_i}
P_{\rm orb}^{-0.20\left(b_i+4a_i/3\right)}.
\label{eq:generic_Delta_numeric}
\end{equation}
The right-hand side is the numerical form used below.
The $-2.00$ term is the adopted $T_{\rm int}^{4}$ cooling penalty.

\subsection{Active-heating channels}\label{subsec:active_radius_scalings}

\subsubsection{Ohmic dissipation}
Following the order-of-magnitude atmospheric induction model of \citet{Menou12}, for which $Q_{\rm ohm}=V_{\phi}^{2}B^{2}/(4\pi\eta)$ and $P_{\rm ohm}=4\pi R_0^{2}\int Q_{\rm ohm}H_p,d\ln P$, and using $H_p\propto T/g$ with $g\propto M_p/R_0^{2}$, we approximate the remaining irradiation dependence locally as $F_{\rm inc}^{\beta_{\Omega}}$ and write
\begin{equation}
P_{\Omega}\propto
B^2R_0^4M_p^{-1}F_{\rm inc}^{\beta_{\Omega}}
\propto
B^2R_0^4M_p^{-1}
T_{\rm eff}^{4\beta_{\Omega}}
P_{\rm orb}^{-4\beta_{\Omega}/3}.
\label{eq:Pohm_numeric}
\end{equation}
Menou-type scalings imply that ohmic power rises faster than the absorbed stellar power on the cool side of the inflation sequence, while magnetic drag produces a turnover along the equivalent equilibrium-temperature coordinate near $T_{\rm eq}\simeq1500$ to $2000$\,K \citep{Menou12,Rogers14,Thorngren18}.
In the final variables, this turnover is a turnover in the irradiation combination $T_{\rm eff}P_{\rm orb}^{-1/3}$.
We adopt a representative pre-turnover local slope
\begin{equation}
\beta_{\Omega}=1.25,
\end{equation}
which corresponds to $(s,u,a,b)=(4,-1,1.25,0)$ in Eq.~(\ref{eq:generic_Delta_numeric}).
The resulting scaling is
\begin{equation}
\boxed{
\Delta_{\Omega}\propto
M_p^{-0.62}\,
P_{\rm orb}^{-0.33}\,
T_{\rm eff}^{1.00}}
\label{eq:Delta_ohmic_numeric}
\end{equation}
for fixed $B$.
At the empirical efficiency peak, the effective $\beta_{\Omega}$ is closer to unity, which gives
\begin{equation}
\Delta_{\Omega,\,\beta_{\Omega}=1}
\propto
M_p^{-0.62}\,
P_{\rm orb}^{-0.27}\,
T_{\rm eff}^{0.80}.
\label{eq:Delta_ohmic_peak_numeric}
\end{equation}

\subsubsection{Thermal tides}
For the Gold-Soter thermal-tide estimate,
\begin{equation}
P_{\rm th,GS}
\propto R_0^4P_{\rm orb}^{-2}T_{\rm eq}^{3}
\propto R_0^4T_{\rm eff}^{3}P_{\rm orb}^{-3}
\propto R_0^4P_{\rm orb}^{-2}F_{\rm inc}^{3/4}
\label{eq:Pthermal_GS_numeric}
\end{equation}
\citep{Socrates13}.
Thus $(s,u,a,b)=(4,0,3/4,2)$ and
\begin{equation}
\boxed{
\Delta_{\rm th,GS}\propto
M_p^{-0.42}\,
P_{\rm orb}^{-0.60}\,
T_{\rm eff}^{0.60}}.
\label{eq:Delta_thermal_GS_numeric}
\end{equation}
The dynamical thermal-tide estimate instead has
\begin{equation}
P_{\rm th,dyn}\propto R_0^5P_{\rm orb}^{-4}
\label{eq:Pthermal_dyn_numeric}
\end{equation}
at fixed $Q'_p$ and a normalized forcing frequency.
With $(s,u,a,b)=(5,0,0,4)$,
\begin{equation}
\boxed{
\Delta_{\rm th,dyn}\propto
M_p^{-0.44}\,
P_{\rm orb}^{-0.80}\,
T_{\rm eff}^{0}}.
\label{eq:Delta_thermal_dyn_numeric}
\end{equation}
If the forcing-frequency normalization is assumed to vary with the mean motion, the period exponent becomes even steeper.

\subsubsection{Kinetic or mechanical heat burial}
For mechanical heating or heat burial, we use
\begin{equation}
P_{\rm kin}=\epsilon_{\rm kin}\pi R_0^2F_{\rm inc}
\propto
\epsilon_{\rm kin}R_0^2T_{\rm eff}^{4}P_{\rm orb}^{-4/3}.
\label{eq:Pkin_numeric}
\end{equation}
Youdin and Mitchell's mechanical-greenhouse model does not require a universal power-law increase of $\epsilon_{\rm kin}$ with irradiation, so we use the constant-efficiency closure $\epsilon_{\rm kin}={\rm const.}$ \citep{Youdin10}.
This gives $(s,u,a,b)=(2,0,1,0)$ and
\begin{equation}
\boxed{
\Delta_{\rm kin}\propto
M_p^{-0.40}\,
P_{\rm orb}^{-0.27}\,
T_{\rm eff}^{0.80}}.
\label{eq:Delta_kinetic_numeric}
\end{equation}

\subsubsection{Maintained eccentricity or obliquity tides}
For eccentricity tides in a synchronously rotating planet,
\begin{equation}
P_{\rm tide}\propto
R_0^5\left(\frac{e^2}{Q'_p}\right)P_{\rm orb}^{-5}
\label{eq:Ptide_numeric}
\end{equation}
at fixed stellar mass.
Obliquity tides have the same role after replacing $e^2/Q'_p$ by the corresponding obliquity-damping factor.
With $(s,u,a,b)=(5,0,0,5)$,
\begin{equation}
\boxed{
\Delta_{\rm tide}\propto
M_p^{-0.44}\,
P_{\rm orb}^{-1.00}\,
T_{\rm eff}^{0}}
\label{eq:Delta_tides_numeric}
\end{equation}
for fixed $e$ or obliquity and fixed $Q'_p$.

\subsection{Delayed-cooling and circulation channels}\label{subsec:delayed_radius_scalings}

Delayed-cooling mechanisms do not correspond to a separate additive $P_i$ and therefore are not described by Eq.~(\ref{eq:generic_Delta_numeric}).
For the pure radiative-blanketing or opacity-delayed-cooling limit, analytic irradiated-envelope models give a weak dependence of the interior entropy on irradiation \citep{Burrows07,Ginzburg15}.
At fixed age, the passive irradiated-envelope cooling relation gives approximately
\begin{equation}
T_c\propto M_p^{1/4}R_0^{-1/2}T_{\rm eq}^{1/4},
\label{eq:passive_Tc_scaling}
\end{equation}
and the small-inflation radius response gives
\begin{equation}
\frac{\Delta R}{R_0}\propto
\frac{T_c}{gR_0}
\propto
\frac{T_cR_0}{M_p}.
\label{eq:passive_radius_response}
\end{equation}
Using $R_0\propto M_p^{-0.06}$ and $T_{\rm eq}\propto T_{\rm eff}P_{\rm orb}^{-1/3}$ gives
\begin{equation}
\boxed{
\Delta_{\rm rad}\propto
M_p^{-0.78}\,
P_{\rm orb}^{-0.08}\,
T_{\rm eff}^{0.25}}.
\label{eq:Delta_radiative_numeric}
\end{equation}
Equation~(\ref{eq:Delta_radiative_numeric}) should be interpreted as the passive radiative-blanketing limit.
It produces a much weaker period and stellar-effective-temperature dependence than the active heating channels.

For layered convection with a fixed compositional-staircase strength, the leading dependence is through surface gravity rather than irradiation.
Approximating the imposed entropy perturbation as fixed gives
\begin{equation}
\boxed{
\Delta_{\rm layer}\propto
M_p^{-1.06}\,
P_{\rm orb}^{0}\,
T_{\rm eff}^{0}}.
\label{eq:Delta_layer_numeric}
\end{equation}
Thus layered convection can change the normalization of $R_p$ substantially, but by itself it does not predict a direct period or stellar-effective-temperature edge unless the layer strength correlates with irradiation.

For potential-temperature advection, circulation models show that non-uniform irradiation can drive the deep atmosphere toward a hotter adiabat than in one-dimensional radiative-convective models \citep{Tremblin17,Sainsbury19}.
Those simulations do not provide a unique universal power law, so we close the analytic scaling with the simplest irradiation-set deep-adiabat assumption, $\delta T_{\rm deep}\propto T_{\rm eq}$.
Using $\Delta R/R_0\propto\delta T_{\rm deep}/(gR_0)$ gives
\begin{equation}
\boxed{
\Delta_{\rm adv}\propto
M_p^{-1.06}\,
P_{\rm orb}^{-0.33}\,
T_{\rm eff}^{1.00}}.
\label{eq:Delta_advection_numeric}
\end{equation}
The mass exponent is steep because a fixed deep-temperature perturbation produces a smaller fractional radius response at higher surface gravity.

\section{Detailed Methods}\label{app:methods_details}
\subsection{Dataset}

Our dataset was constructed from the \texttt{Encyclopaedia of exoplanetary systems} (\url{https://exoplanet.eu/catalog/}) by applying uniform cuts in mass, radius, and orbital period.
Specifically, we retained only planets with cataloged masses in the range \(0.5\,\mathrm{M}_{\jupiter} \leq M_{\rm p} \leq 2.0\,\mathrm{M}_{\jupiter}\), orbital periods \(P_{\rm orb} \leq 10\)\,days, and radii satisfying \(0.5\,\mathrm{R}_{\jupiter} < R_{\rm p} \leq 3.0\,\mathrm{R}_{\jupiter}\), for 328 data points in total (Fig.~\ref{fig:pair-plot}).
These criteria select short-period, Jupiter-scale exoplanets while excluding objects with very small radii, extremely inflated radii, or masses outside the adopted giant-planet regime.
Because the selection was implemented through direct numerical comparisons, planets lacking reported values in any of the filtered quantities were not included in the final sample.
For the control sample, we constructed a super-Earth data set from the same catalog using \(R_p \leq 1.8\,R_\oplus\) and requiring complete measurements of \(M_p\), \(R_p\), \(P_{\rm orb}\), and \(T_{\rm eff}\).
This yielded 179 exoplanets (Fig. \ref{fig:pair-plot2}). 
No additional causal-discovery variables were used in the control analysis.

\subsection{Causal discovery}

We inferred the causal structure among planetary mass $M_{\rm p}$, planetary radius $R_{\rm p}$, orbital period $P_{\rm orb}$, and stellar effective temperature $T_{\rm eff}$ using Best Order Score Search \citep[BOSS;][]{andrews2023fast} with the Fourier Feature Marginal Likelihood (FFML) score \citep{ramsey2026fourier}.

\subsubsection{Best Order Score Search (BOSS)}
BOSS is a score-based method for learning directed acyclic graphs (DAGs). 
The original BOSS algorithm was described with a BIC local score; here we use the same search procedure, but replace the local score by FFML to allow nonlinear conditional relationships.

Let $X_1,\ldots,X_p$ denote the variables, with $p=4$ in the main analysis. 
For a DAG $G$, let $\operatorname{Pa}_G(j)$ be the parent set of $X_j$. 
The score of $G$ is decomposable:
\begin{equation}
S(G)
=
\sum_{j=1}^{p}
S_{\rm FFML}
\left(
X_j \mid X_{\operatorname{Pa}_G(j)}
\right).
\label{eq:boss_graph_score}
\end{equation}
Thus each candidate edge is evaluated by asking whether it improves the conditional model of the child after accounting for its other parents.

BOSS searches over variable orderings rather than directly enumerating all DAGs. 
For an ordering $\pi$, define the prefix of $X_j$ as
\begin{equation}
\operatorname{Pre}_{\pi}(j)
=
\{k:\pi(k)<\pi(j)\}.
\label{eq:boss_prefix}
\end{equation}
Only variables in $\operatorname{Pre}_{\pi}(j)$ are allowed to be parents of $X_j$, which guarantees that the resulting graph is acyclic. 
For a fixed ordering, BOSS projects the ordering to a DAG by selecting parents for each node using a grow--shrink procedure. 
Starting from an empty parent set $A=\emptyset$, the grow step repeatedly adds the prefix variable that gives the largest positive gain
\begin{equation}
\Delta_{\rm add}(k;A,j)
=
S_{\rm FFML}(X_j\mid X_{A\cup\{k\}})
-
S_{\rm FFML}(X_j\mid X_A),
\label{eq:boss_grow_gain}
\end{equation}
until no addition improves the score. 
The shrink step then checks the selected parents and removes any parent whose deletion gives a positive gain
\begin{equation}
\Delta_{\rm del}(k;A,j)
=
S_{\rm FFML}(X_j\mid X_{A\setminus\{k\}})
-
S_{\rm FFML}(X_j\mid X_A).
\label{eq:boss_shrink_gain}
\end{equation}
The final parent set is denoted $\widehat{\operatorname{Pa}}_{\pi}(j)$, and the projected DAG has edges $X_k\rightarrow X_j$ for all $k\in\widehat{\operatorname{Pa}}_{\pi}(j)$.

BOSS then improves the ordering itself. 
For each variable, it tries moving that variable to each possible position in the ordering, recomputes the projected-DAG score, and keeps the move only if it increases the total score in Eq.~(\ref{eq:boss_graph_score}). 
This sweep over variables is repeated until no single-variable move improves the score. 
Grow-shrink trees cache the local parent-set scores encountered during this process, so parent sets shared by nearby orderings are not recomputed. 
After the best ordering is found, BOSS converts the selected DAG into a partially-directed graph when orientations are not identifiable from the score and conditional independence structure alone. 
When FFML is used in DAG mode, it can also prefer one direction within a Markov-equivalent pair because FFML is not score equivalent.

\subsubsection{Search criteria: Fourier Feature Marginal Likelihood (FFML) score}

The FFML score is a nonlinear marginal-likelihood score. 
For each local regression, let $Y\in\mathbb{R}^n$ be the child variable observed for $n$ planets, and let $Z\in\mathbb{R}^{n\times d}$ be a candidate parent set. 
FFML starts from the Gaussian-process model
\begin{equation}
Y=f(Z)+\epsilon,
\qquad
\epsilon\sim\mathcal{N}(0,\sigma^2 I_n),
\qquad
f\sim\mathcal{GP}(0,k),
\label{eq:ffml_gp_model}
\end{equation}
where $k$ is an RBF kernel. 
Let $K_Z$ be the kernel matrix over the parent observations and define
\begin{equation}
C_Z=K_Z+\sigma^2 I_n.
\label{eq:ffml_covariance}
\end{equation}
After integrating out the unknown function $f$, the exact kernel marginal log-likelihood, up to constants independent of the parent set, is
\begin{equation}
S_{\rm KML}(Y\mid Z)
=
-\frac{1}{2}Y^{\top}C_Z^{-1}Y
-\frac{1}{2}\log |C_Z|.
\label{eq:kml_score}
\end{equation}
The first term rewards fit, while the log-determinant term is an Occam factor that penalizes overly flexible parent sets.

Direct evaluation of Eq.~(\ref{eq:kml_score}) requires operations on an $n\times n$ matrix. 
FFML avoids this by approximating the RBF kernel with random Fourier features. 
For
\begin{equation}
k(z,z')
=
\exp\left(-\frac{\|z-z'\|^2}{h^2}\right),
\label{eq:rbf_kernel}
\end{equation}
draw frequencies $\omega_{\ell}\sim\mathcal{N}(0,2h^{-2}I_d)$ and phases $b_{\ell}\sim\operatorname{Unif}(0,2\pi)$, and define
\begin{equation}
\phi(z)
=
\sqrt{\frac{2}{m}}
\begin{bmatrix}
\cos(\omega_1^{\top}z+b_1)\\
\vdots\\
\cos(\omega_m^{\top}z+b_m)
\end{bmatrix}.
\label{eq:rff_map}
\end{equation}
Stacking $\phi(z_i)$ over all planets gives $\Phi\in\mathbb{R}^{n\times m}$, with
\begin{equation}
K_Z\approx \Phi\Phi^{\top}.
\label{eq:rff_kernel_approx}
\end{equation}
Define
\begin{equation}
H=\Phi^{\top}\Phi,
\qquad
r=\Phi^{\top}Y.
\label{eq:ffml_H_r}
\end{equation}
Using the Woodbury identity and the matrix determinant lemma, FFML computes the local score as
\begin{equation}
\begin{aligned}
S_{\rm FFML}(Y \mid Z)
= -\frac{1}{2\sigma^2}
\left[
Y^\top Y
-
r^\top \left(H+\sigma^2 I_m\right)^{-1} r
\right]
-\frac{1}{2}
\left[
(n-m)\log \sigma^2
+
\log \left|H+\sigma^2 I_m\right|
\right].
\label{eq:ffml_score}
\end{aligned}
\end{equation}
All linear solves and log determinants are now evaluated in the \(m\)-dimensional feature space rather than the \(n\)-dimensional sample space.
For one local FFML score with \(d\) parents and \(m\) random features, forming the feature matrix \(\Phi\) costs \(\mathcal{O}(ndm)\), forming \(H=\Phi^\top\Phi\) costs \(\mathcal{O}(nm^2)\), and factorizing \(H+\sigma^2 I_m\) costs \(\mathcal{O}(m^3)\).
Thus the per-parent-set score cost is \(\mathcal{O}(ndm+nm^2+m^3)\), ignoring lower-order terms.
For fixed \(d\) and \(m\ll n\), this score evaluation is effectively linear in the number of planets.
The total BOSS runtime further depends on the number of order moves and candidate parent sets evaluated.
When the parent set is empty, FFML reduces to the Gaussian noise model $Y\sim\mathcal{N}(0,\sigma^2 I_n)$.

All variables in our analysis are continuous, so the mixed discrete-parent extension of FFML is not needed. 
Continuous variables are standardized for kernel evaluation, and the RBF bandwidth is selected from pairwise distances among candidate parent observations.
The \(\mathcal{O}(ndm+nm^2+m^3)\) scaling refers to the FFML linear-algebra evaluation for a fixed bandwidth and random-feature basis.
A naive all-pairs bandwidth selection for each candidate parent set would add \(\mathcal{O}(n^2d)\) to the corresponding local-score evaluation.
The random feature basis is coupled by target variable, so local-score differences such as $S_{\rm FFML}(Y\mid Z\cup\{X\})-S_{\rm FFML}(Y\mid Z)$ are evaluated with the same random features and are therefore more stable.

\bibliography{bibliography}{}

\end{document}